\pdfoutput=1
\documentclass{article}
\usepackage[turkish]{babel}
\usepackage[T1]{fontenc}
\usepackage{graphicx}
\usepackage[pdftex]{hyperref}
\usepackage{authblk}

\title{Ankara \"{U}niversitesi Kreiken Rasathanesinde Bulunan Radyo Teleskobun Y\"{o}nlendirme Mekanizmas{\i}n{\i}n \.{I}ncelenip Yenilenmesi, Anten Benzetimleri ve Uyumlama Tasar{\i}m{\i}}

\author[1,2]{T\"{u}rker Dolap\c{c}{\i}}

\author[2]{\c{S}ahin K\"{u}li\u{g}}

\author[1]{\"{O}zg\"{u}r Erg\"{u}l}

\affil[1]{Orta Do\u{g}u Teknik \"{U}niversitesi \par Elektrik - Elektronik M\"{u}hendisli\u{g}i B\"{o}l\"{u}m\"{u} \par Ankara \par \href{mailto:dolapci.turker@metu.edu.tr}{dolapci.turker@metu.edu.tr}, \href{mailto:ozergul@metu.edu.tr}{ozergul@metu.edu.tr}}

\affil[2]{T\"{u}rkiye Amat\"{o}r Uydu Teknolojileri Derne\u{g}i (TAMSAT) \par Ankara \par \href{mailto:sahin.kulig@gmail.com}{sahin.kulig@gmail.com}}

\begin{document}
	
	\maketitle
	
	\begin{center}
		\textbf{\"{O}zet}
	\end{center}
	\par
	A.\"{U}. Kreiken Rasathanesinde bulunan, tam dalgaboyunda dipol elemanlara sahip VHF ve UHF anten dizilerinden olu\c{s}an tarihi radyo teleskobun faal hale getirilmesi i\c{c}in anten y\"{o}nlendirme mekanizmas{\i} incelenip tamir edilmi\c{s}, teleskop anteninin uyumlama tasar{\i}m{\i} benzetim ortam{\i}nda ger\c{c}ekle\c{s}tirilmi\c{s}tir. K{\i}sa devre ile sonland{\i}r{\i}lm{\i}\c{s} paralel iletim hatt{\i} saplamalar{\i} kullanarak uyumlama tasarlanm{\i}\c{s}t{\i}r. Dipol anten elemanlar{\i}n{\i}n boyunun yayg{\i}n tercih edilen yar{\i}m dalgaboyu yerine y\"{u}ksek kazan\c{c}l{\i} tam dalgaboyu olmas{\i}, antenin ve uyumlama y\"{o}nteminin \"{o}nemini art{\i}rmaktad{\i}r.
	\par
	\medskip 
	\begin{center}
		\textbf{Abstract}
	\end{center}
	\par
	To make the historical radio telescope of A.\"{U}. Kreiken Observatory that includes full-wavelength VHF and UHF dipole arrays operational, the antenna rotator mechanism is renovated and the antenna matching structure is designed in a simulation environment. The matching structure is designed by using short-circuited parallel transmission line stubs. The use of high-gain full-wavelength elements, as opposed to the conventional half-wavelength elements, increases the importance of the structure and the matching method.
	
	\section{Giri\c{s}}
	Ankara \"{U}niversitesi Kreiken Rasathanesinde \cite{cite_key1} bulunan, G\"{u}ne\c{s} leke g\"{o}zlemlerinde kullan{\i}lan radyo teleskop, d\"{u}zlemsel yans{\i}t{\i}c{\i} \"{u}zerine konumland{\i}r{\i}lm{\i}\c{s} $200$ MHz ve $545$ MHz'te \c{c}al{\i}\c{s}an tam-dalga dipol anten dizileri, bu anten dizilerini d\"{o}nd\"{u}rmeye yarayan motorlu bir y\"{o}nlendirme mekanizmas{\i}, ve bir alma\c{c}tan olu\c{s}maktad{\i}r. 1960--1980'li y{\i}llarda kullan{\i}lan bu radyo teleskop g\"{u}n\"{u}m\"{u}zde faal de\u{g}ildir. Sistemin, g\"{u}n\"{u}m\"{u}z teknolojisinden bir hayli geride olmas{\i}na kar\c{s}{\i}n, tarihi de\u{g}er ta\c{s}{\i}mas{\i} ve dayan{\i}kl{\i} yap{\i}s{\i} sayesinde hala fiziksel olarak iyi durumda olmas{\i} yenilenme ihtimalini g\"{u}ndeme getirmi\c{s}tir. Bu do\u{g}rultuda, anten y\"{o}nlendirme mekanizmas{\i} incelenmi\c{s} ve \c{c}al{\i}\c{s}{\i}r hale getirilmi\c{s}tir. Anten benzetimleri MoM y\"{o}ntemini kullanan NEC tabanl{\i} 4NEC2 \cite{cite_key2} program{\i}yla ger\c{c}ekle\c{s}tirilmi\c{s}tir. Benzetimler vas{\i}tas{\i}yla kazan\c{c} analizlerinin yan{\i} s{\i}ra, \c{c}e\c{s}itli empedanslara sahip iletim hatlar{\i}n{\i}n kullan{\i}lmas{\i}yla anten dizilerinin $50$ Ohm'a uyumlamas{\i} sa\u{g}lanm{\i}\c{s}t{\i}r. Tam dalgaboyundaki elemanlara sahip bir dipol anten dizisinin \c{c}eyrek dalga boyu d\"{o}n\"{u}\c{s}t\"{u}r\"{u}c\"{u}ler ve k{\i}sa devre yap{\i}lm{\i}\c{s} saplamalarla uyumlanmas{\i}, literat\"{u}rde kar\c{s}{\i}la\c{s}{\i}lmam{\i}\c{s} \"{o}zg\"{u}n bir \c{c}al{\i}\c{s}ma olarak \"{o}ne \c{c}{\i}kmaktad{\i}r.
	
	\section{Y\"{o}nlendirme Mekanizmas{\i}}
	Y\"{o}nlendirme mekanizmas{\i}, anten dizilerinin G\"{u}ne\c{s}'i takip etmesi i\c{c}in motorlar arac{\i}l{\i}\u{g}{\i} ile yanca ekseninde hareket sa\u{g}lamaktad{\i}r. Mevsime ba\u{g}l{\i} olarak de\u{g}i\c{s}en G\"{u}ne\c{s} pozisyonuna g\"{o}re ayarlanmas{\i} gereken y\"{u}kseli\c{s} a\c{c}{\i}s{\i} ise bir el mekanizmas{\i} ile de\u{g}i\c{s}tirilmektedir. \c{C}al{\i}\c{s}ma ba\c{s}lang{\i}c{\i}nda yap{\i}lan ilk incelemelerde, y\"{o}nlendirme mekanizmas{\i}n{\i}n uzun s\"{u}redir kullan{\i}lmamas{\i}ndan dolay{\i} di\c{s}lilerinin aras{\i}nda bulunan gres ya\u{g}{\i}n{\i}n kat{\i}la\c{s}t{\i}\u{g}{\i} ve bu y\"{u}zden \c{c}al{\i}\c{s}mad{\i}\u{g}{\i} g\"{o}zlemlenmi\c{s}tir. \.{I}lk olarak, t\"{u}m aksam ya\u{g} \c{c}\"{o}z\"{u}c\"{u} ile temizlenmi\c{s} ve a\u{g}{\i}r \c{c}evresel ko\c{s}ullarda kullan{\i}ma uygun tipte gres ya\u{g}{\i} ile ya\u{g}lanm{\i}\c{s}t{\i}r. Bu sayede, y\"{o}nlendirme mekanizmas{\i}n{\i}n hareket etme yetene\u{g}ini tekrar kazanmas{\i} sa\u{g}lanm{\i}\c{s}t{\i}r.
	
	\c{S}ehir \c{s}ebekesi elektri\u{g}iyle \c{c}al{\i}\c{s}t{\i}r{\i}lan motorlardaki elektriksel donan{\i}m ba\c{s}tan sonra kontrol edilmi\c{s}, mekanizma par\c{c}alar{\i}n{\i}n fonksiyonlar{\i} incelenmi\c{s}tir. Anten \"{u}zerindeki iki adet tek fazl{\i} asenkron motordan birinin G\"{u}ne\c{s} takibi i\c{c}in hareketi, di\u{g}erinin ise h{\i}zl{\i} ba\c{s}a almay{\i} sa\u{g}lad{\i}\u{g}{\i} g\"{o}zlemlenmi\c{s}tir. G\"{u}ne\c{s} takibini sa\u{g}layan motorun, di\c{s}lilerin de yard{\i}m{\i}yla G\"{u}ne\c{s}'in hareket h{\i}z{\i}na uyumlu hareket ger\c{c}ekle\c{s}tirdi\u{g}i anla\c{s}{\i}lm{\i}\c{s}t{\i}r. Asenkron motorlar ilk kalk{\i}\c{s} an{\i}nda harekete ge\c{c}me e\u{g}iliminde de\u{g}ildir. Hareketi ba\c{s}latmak i\c{c}in, ayn{\i} zamanda hareket y\"{o}n\"{u}n\"{u} de belirleyen, yard{\i}mc{\i} sarg{\i} i\c{c}ermektedirler. Hareketi s\"{u}rd\"{u}ren ana sarg{\i}ya s\"{u}rekli, hareketi ba\c{s}latan yard{\i}mc{\i} sarg{\i}ya ise k{\i}sa bir s\"{u}re enerji verilmesi gerekmektedir. G\"{u}ne\c{s} takibi i\c{c}in hareketi sa\u{g}layan motorun yard{\i}mc{\i} sarg{\i}lar{\i}na enerji gitmedi\u{g}i ve ba\c{s}lang{\i}\c{c} r\"{o}lesinin ar{\i}zal{\i} oldu\u{g}u, bu y\"{u}zden d\"{o}n\"{u}\c{s} hareketinin ba\c{s}layamad{\i}\u{g}{\i} tespit edilmi\c{s}tir. Yakla\c{s}{\i}k 70 ya\c{s}{\i}ndaki ar{\i}zal{\i} elektromekanik r\"{o}lenin yenisini bulmak g\"{u}\c{c} olaca\u{g}{\i}ndan, ar{\i}zay{\i} gidermek i\c{c}in ayn{\i} i\c{s}levi yerine getirebilecek g\"{u}ncel teknolojiye sahip elektronik zaman r\"{o}lesi kullan{\i}lm{\i}\c{s}t{\i}r. Yap{\i}lan testlerde yard{\i}mc{\i} sarg{\i}ya yakla\c{s}{\i}k $2$ sn enerji verildi\u{g}inde motorun d\"{o}nmeye ba\c{s}lad{\i}\u{g}{\i} g\"{o}zlemlenmi\c{s}tir. 
	
	Sonu\c{c} olarak, y\"{o}nlendirme mekanizmas{\i} yap{\i}lan bak{\i}m ve tamirattan sonra mekanik ve elektriksel olarak \c{c}al{\i}\c{s}{\i}r hale getirilmi\c{s}tir.
	
	\section{Anten Dizileri}
	Y\"{o}nlendirme mekanizmas{\i}n{\i}n \"{u}zerinde $200$ MHz'te \c{c}al{\i}\c{s}an yatay polarizasyona sahip iki elemanl{\i} ve $545$ MHz'te \c{c}al{\i}\c{s}an dikey polarizasyona sahip sekiz elemanl{\i} anten dizileri bulunmaktad{\i}r. Dizi elemanlar{\i} olarak \c{c}al{\i}\c{s}ma frekanslar{\i}nda tam dalgaboyu uzunlu\u{g}una sahip dipol antenler kullan{\i}lm{\i}\c{s}t{\i}r. Anten elemanlar{\i}, \"{u}zerinde bulunduklar{\i} yans{\i}t{\i}c{\i} {\i}zgaradan \c{c}eyrek dalgaboyu kadar yukar{\i}ya yerle\c{s}tirilmi\c{s}tir. Boyutlar{\i} $215 \times 135$ cm olan yans{\i}t{\i}c{\i} {\i}zgaray{\i} olu\c{s}turan \c{c}ubuklar, \c{c}al{\i}\c{s}ma frekanslar{\i}ndaki dalgaboylar{\i}n{\i}n onda birine denk gelen aral{\i}klarla dizilmi\c{s}tir. Yatay antenlerin \c{c}ap{\i} $1.5$ cm, dikey antenlerin \c{c}ap{\i} $1$ cm, {\i}zgaray{\i} olu\c{s}turan \c{c}ubuklar{\i}n \c{c}ap{\i} $1$ cm'dir. 
	
\begin{figure}[b!]
		\centering
		\shorthandoff{=}
		\includegraphics[trim=0 0 0 0, clip,width=\textwidth]{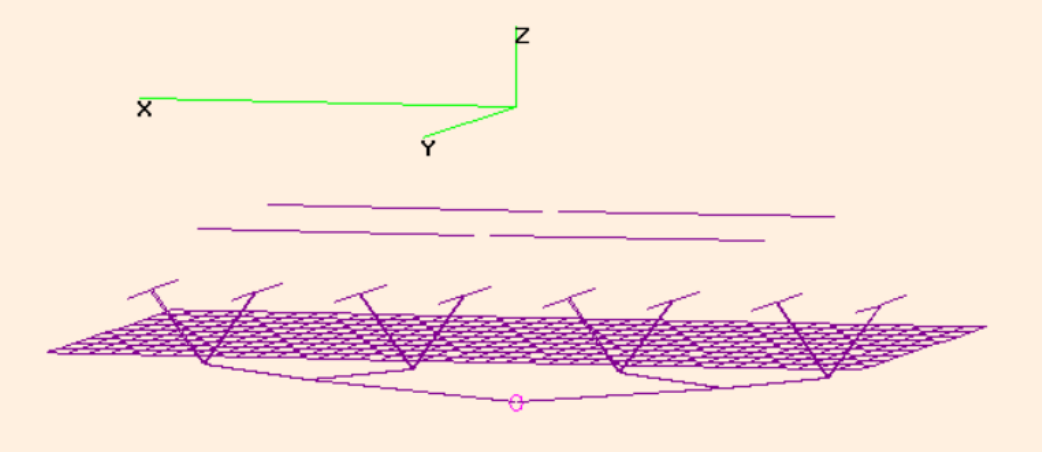}
		\shorthandon{=}
		\caption{Benzetim ortam{\i}nda olu\c{s}turulan anten modeli.}
		\label{fig:antenmodel}
	\end{figure}

	\c{C}al{\i}\c{s}ma frekans{\i} $545$ MHz olan dikey dipol dizisinin elemanlar{\i} aras{\i}ndaki ba\u{g}lant{\i}lar, aralar{\i}nda $1.5$ cm uzakl{\i}k bulunan $0.5$ cm \c{c}ap{\i}ndaki paralel iletkenli dengeli iletim hatlar{\i}yla sa\u{g}lanm{\i}\c{s}t{\i}r. Dizi elemanlar{\i} ikinin kuvvetleri olarak \"{u}\c{c} a\c{s}amada birle\c{s}tirilmi\c{s}tir. Antenlerden ortak besleme noktas{\i}na do\u{g}ru uzanan ilk grup iletim hatt{\i} yar{\i}m dalgaboyu, ikinci grup iletim hatt{\i} yar{\i}m dalgaboyu, \"{u}\c{c}\"{u}nc\"{u} grup (ortak besleme noktas{\i}na en yak{\i}n) iletim hatt{\i} ise tam dalgaboyu uzunlu\u{g}undad{\i}r. \c{C}al{\i}\c{s}ma frekans{\i} $200$ MHz olan yatay dipol dizisinde ise herhangi bir besleme yap{\i}s{\i} mevcut de\u{g}ildir. Bu anlat{\i}lan duruma g\"{o}re 4NEC2 program{\i} vas{\i}tas{\i}yla olu\c{s}turulan model \c{S}ekil \ref{fig:antenmodel}'de g\"{o}sterilmi\c{s}tir.
	
	Rasathanede bulunan yaz{\i}l{\i} kaynaklarda \cite{cite_key3}, $545$ MHz frekans{\i}ndaki dizinin ortaklanm{\i}\c{s} ba\u{g}lant{\i} yerinin uygun bir balun ve $50$ Ohm empedans de\u{g}erine sahip e\c{s}eksenli bir iletim hatt{\i} vas{\i}tas{\i}yla almaca ba\u{g}land{\i}\u{g}{\i} belirtilmi\c{s}tir. Di\u{g}er ($200$ MHz) frekanstaki dizinin elemanlar{\i}n{\i}n ise, \c{c}eyrek dalgaboyu uzunlu\u{g}undaki iletim hatlar{\i}yla birbirlerine ba\u{g}land{\i}klar{\i}, sonras{\i}nda uygun bir balun ve $50$ Ohm empedans de\u{g}erine sahip e\c{s}eksenli bir iletim hatt{\i} ile almaca ba\u{g}land{\i}klar{\i} belirtilmi\c{s}tir. 
	
	Tam dalgaboyu dipoller genelde geometrik merkezleri d{\i}\c{s}{\i}ndaki bir veya birden fazla noktadan beslenerek kullan{\i}l{\i}rlar. Bunun nedeni orta noktadaki ak{\i}m yo\u{g}unlu\u{g}unun d\"{u}\c{s}\"{u}k olmas{\i} ve \c{c}ok y\"{u}ksek empedans de\u{g}eridir \cite{cite_key4}. Merkezden besleme i\c{c}in ise mutlaka uyumlama devresi tasar{\i}m{\i} gerekmektedir \cite{cite_key5}. Birden fazla eleman{\i}n paralel ba\u{g}lanmas{\i}yla elde edilen dizide empedans, tek eleman empedans{\i}na k{\i}yasla d\"{u}\c{s}\"{u}k olsa da, $50$ Ohm'dan y\"{u}ksektir. 
	
\begin{figure}[]
		\centering
		\shorthandoff{=}
		\includegraphics[trim=0 0 0 0, clip,width=0.8\textwidth]{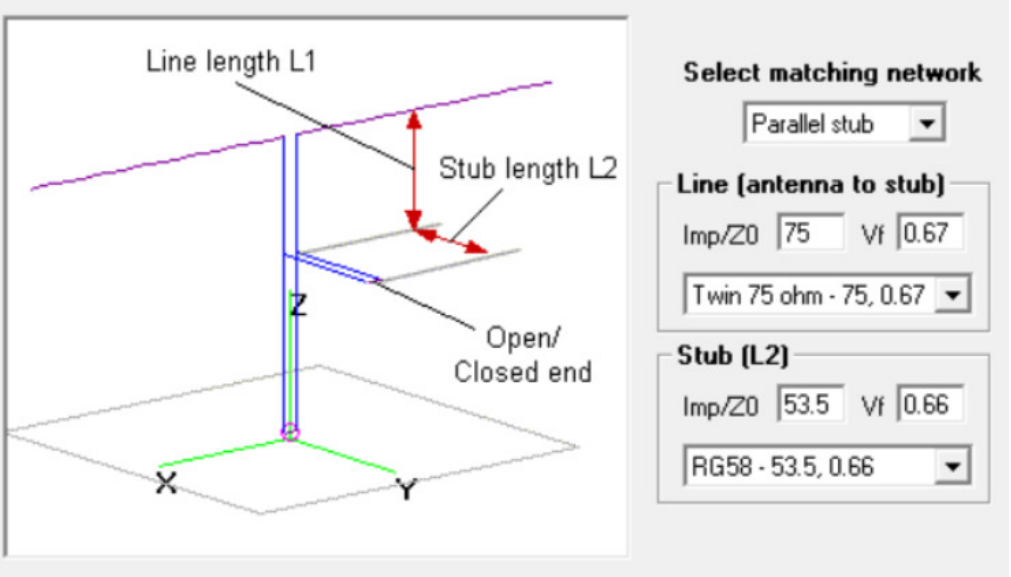}
		\shorthandon{=}
		\caption{4NEC2 program{\i}nda bulunan uyumlama arac{\i}.}
		\label{fig:stub_menu}
	\end{figure}

	\c{C}al{\i}\c{s}ma frekans{\i} $200$ MHz olan dizi elemanlar{\i}, benzetim ortam{\i}nda iki adet \c{c}eyrek dalgaboyu uzunlu\u{g}unda $300$ Ohm empedans de\u{g}erine sahip dengeli iletim hatt{\i}yla ortalar{\i}ndan birbirine ba\u{g}lanm{\i}\c{s}t{\i}r. Elde edilen ba\u{g}lant{\i} noktas{\i}ndaki empedans de\u{g}erinin paralel saplama y\"{o}ntemiyle \cite{cite_key6} $50$ Ohm'a uyumlanabilecek mertebede oldu\u{g}u g\"{o}zlemlenmi\c{s}tir. 4NEC2 program{\i}ndaki uyumlama arac{\i}n{\i}n (\c{S}ekil \ref{fig:stub_menu}) kullan{\i}lmas{\i}yla, bu ortak nokta L1 = $30$ cm uzunlu\u{g}undaki $75$ Ohm empedans de\u{g}erine sahip dengeli iletim hatt{\i} ve k{\i}sa devre ile sonland{\i}r{\i}lm{\i}\c{s} L2 = $14$ cm uzunlu\u{g}undaki $53.5$ Ohm empedans de\u{g}erine sahip RG58 e\c{s}eksenli saplama ile $50$ Ohm'a uyumlanm{\i}\c{s}t{\i}r. \c{C}al{\i}\c{s}ma frekans{\i} $545$ MHz olan dizideki eleman ba\u{g}lant{\i}lar{\i} halihaz{\i}rda bulundu\u{g}u i\c{c}in, do\u{g}rudan 4NEC2 program{\i}ndaki uyumlama arac{\i}n{\i}n kullan{\i}lmas{\i}yla paralel saplama y\"{o}ntemiyle uyumlama ger\c{c}ekle\c{s}tirilmi\c{s}tir. Uyumlama arac{\i}yla yap{\i}lan hesaplamalarda, L1 = $10.5$ cm ve L2 = $11.5$ cm kullan{\i}ld{\i}\u{g}{\i}nda anten uyumlamas{\i}n{\i}n ba\c{s}ar{\i}l{\i} olarak ger\c{c}ekle\c{s}ti\u{g}i g\"{o}zlemlenmi\c{s}tir. 
	
\begin{figure}[]
		\centering
		\shorthandoff{=}
		\includegraphics[trim=0 0 0 0, clip,width=\textwidth]{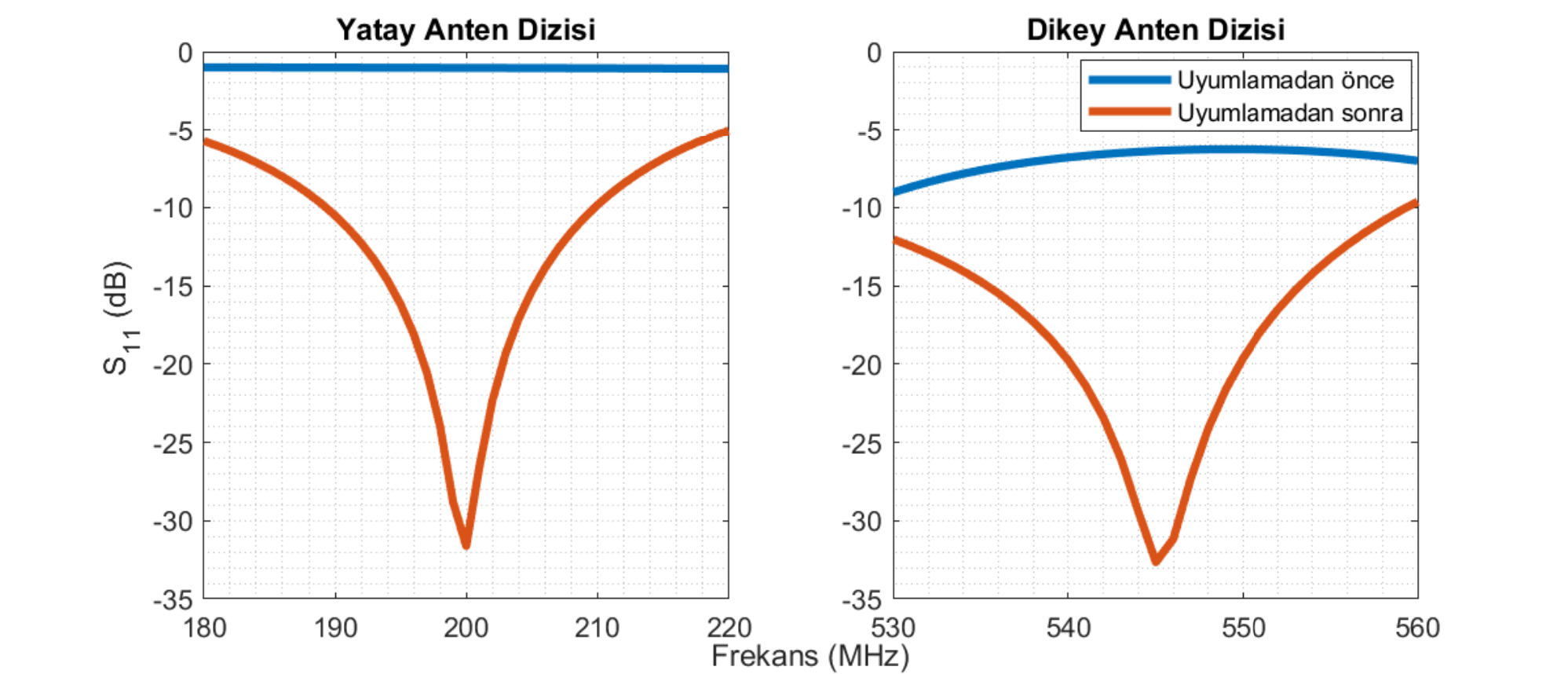}
		\shorthandon{=}
		\caption{Uyumlama \"{o}ncesi ve sonras{\i} yans{\i}ma katsay{\i}lar{\i}.}
		\label{fig:s_param}
	\end{figure}

\begin{figure}[]
		\centering
		\shorthandoff{=}
		\includegraphics[trim=0 0 0 0, clip,width=\textwidth]{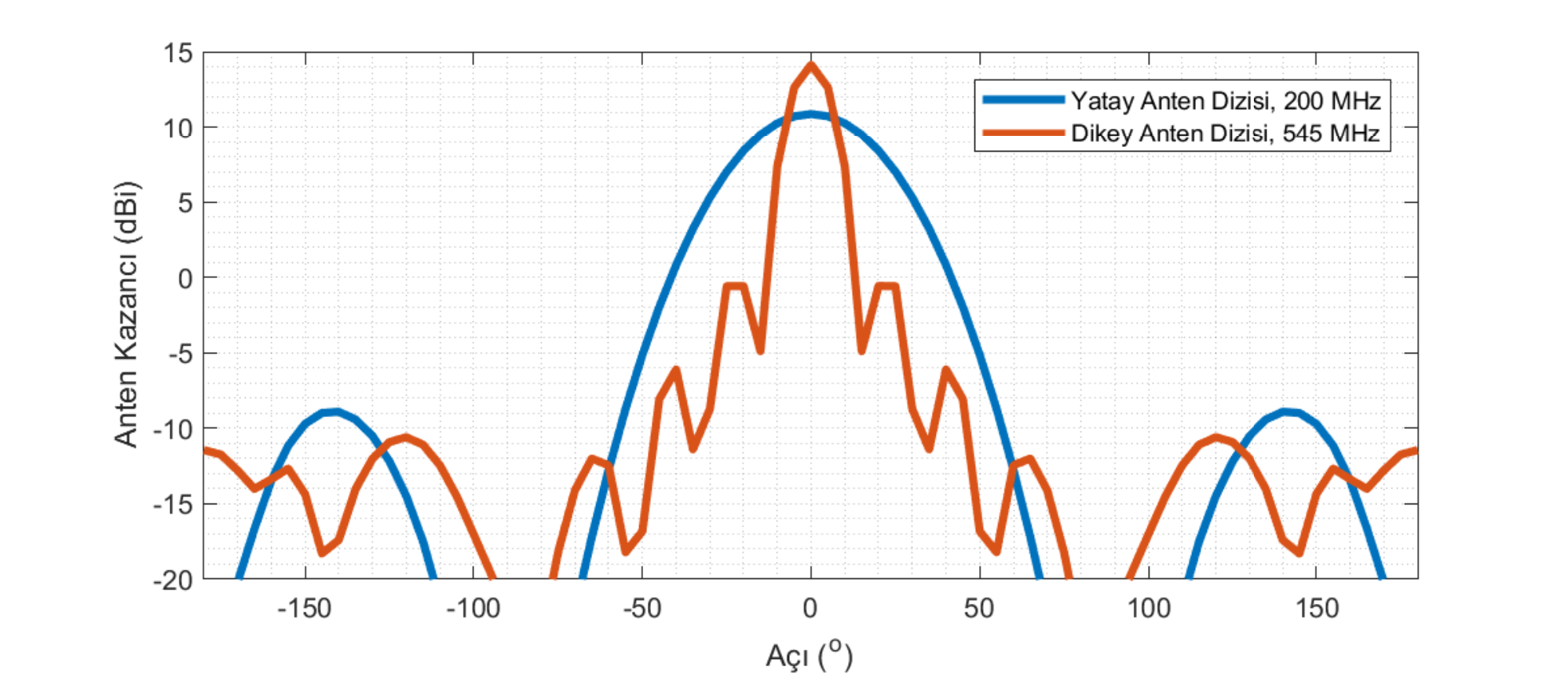}
		\shorthandon{=}
		\caption{Dizilerin merkez \c{c}al{\i}\c{s}ma frekanslar{\i}ndaki kazan\c{c} \"{o}r\"{u}nt\"{u}leri.}
		\label{fig:pattern}
	\end{figure}

	4NEC2 program{\i}yla ger\c{c}ekle\c{s}tirilen uyumlama \"{o}ncesi ve sonras{\i} yans{\i}ma katsay{\i}lar{\i} \c{S}ekil \ref{fig:s_param}'te, dizilerin merkez \c{c}al{\i}\c{s}ma frekanslar{\i}ndaki kazan\c{c} \"{o}r\"{u}nt\"{u}leri ise \c{S}ekil \ref{fig:pattern}'te g\"{o}sterilmi\c{s}tir. Yatay anten dizisinin $3$ dB h\"{u}zme geni\c{s}li\u{g}i $40^\circ$ ve kazanc{\i} $11.5$ dBi, dikey anten dizisinin $3$ dB h\"{u}zme geni\c{s}li\u{g}i $10^\circ$ ve kazanc{\i} $14.2$ dBi olarak hesaplanm{\i}\c{s}t{\i}r. 
	
	\section{Sonu\c{c}}
	Bu \c{c}al{\i}\c{s}mada, A.\"{U}. Kreiken Rasathanesinde bulunan radyo teleskobun y\"{o}nlendirme mekanizmas{\i} ve anten dizileri incelenmi\c{s}tir. Sistemin yeniden faal hale gelmesinde kritik \"{o}neme sahip y\"{o}nlendirme mekanizmas{\i} \c{c}al{\i}\c{s}{\i}r hale getirilmi\c{s}tir. Benzetim ortam{\i}nda ger\c{c}ekle\c{s}tirilen \c{c}al{\i}\c{s}malarla, anten dizilerinin beslemelerinin 50 Ohm'luk alma\c{c}lara uyumlulu\u{g}u sa\u{g}lanm{\i}\c{s}t{\i}r. Tam dalgaboyundaki dipol anten dizilerinin besleme tasar{\i}mlar{\i} \c{c}e\c{s}itli empedanstaki iletim hatlar{\i}n{\i}n kullan{\i}m{\i}yla planlanm{\i}\c{s}, iletim hatt{\i} uzunluklar{\i} 4NEC2 program{\i}ndaki uyumlama arac{\i} vas{\i}tas{\i}yla hesaplanm{\i}\c{s}t{\i}r.
	
	L1 uzunlu\u{g}unda ve $75$ Ohm empedans de\u{g}erine sahip dengeli iletim hatt{\i}ndan sonra kart tipi bir $50$ Ohm $1:1$ balun kullan{\i}lmas{\i}, hemen ard{\i}ndan ise kart \"{u}zerindeki mikro\c{s}erit hatta paralel tak{\i}lacak k{\i}sa devre ile sonland{\i}r{\i}lm{\i}\c{s} RG58 e\c{s}eksenli iletim hatt{\i} ile L2 uzunlu\u{g}unda saplama uyumlamas{\i} ger\c{c}ekle\c{s}tirilmesi, her iki dizi i\c{c}in de kolay ve etkili bir uyumlama ger\c{c}eklemesidir. Benzetim sonucu elde edilen yans{\i}ma katsay{\i}s{\i} de\u{g}erlerinin her iki dizi i\c{c}in de geni\c{s} bir frekans band{\i}nda $-10$ dB seviyesinin alt{\i}nda olmas{\i}, uygun bir uyumlama tasar{\i}m{\i}n{\i}n ger\c{c}ekle\c{s}tirildi\u{g}ini g\"{o}stermektedir.
	
	\section*{Te\c{s}ekk\"{u}r}
	Yazarlar, bu \c{c}al{\i}\c{s}ma fikrini g\"{u}ndeme getiren Bar{\i}\c{s} Din\c{c}'e, \c{c}al{\i}\c{s}man{\i}n yap{\i}lmas{\i} i\c{c}in radyo amat\"{o}rlerine kap{\i}lar{\i}n{\i} a\c{c}an A.\"{U}. Kreiken Rasathanesine, ba\c{s}ta A. Tahir Dengiz olmak \"{u}zere saha \c{c}al{\i}\c{s}mas{\i}nda yer alan TAMSAT \"{u}yelerine, rehberlikleri ve y\"{o}nlendirmeleri i\c{c}in Prof. \"{O}zlem Ayd{\i}n \c{C}ivi, Prof. Sencer Ko\c{c}, ve Prof. Canan Toker'e te\c{s}ekk\"{u}r eder.

\end{document}